\documentclass[aip,preprint]{revtex4-1}
\usepackage{xcolor}
\usepackage{graphicx}
\usepackage{amsmath}
\usepackage{amssymb}
\usepackage{amsfonts}
\usepackage{afterpage}
\newcommand{\diff}[2]%
{
  \frac{d#1}{d#2}
}
\newcommand{\diffs}[2]%
{
  \frac{d^2#1}{d#2^2}
}
\newcommand{\pdiff}[2]%
{
  \frac{\partial#1}{\partial#2}
}
\newcommand{\pdiffs}[2]%
{
  \frac{\partial^2#1}{\partial#2^2}
}
\newcommand{\pdiffx}[3]%
{
  \frac{\partial^2#1}{\partial#2 \partial#3}
}

\newcommand{\tdiff}[2]%
{
  {d#1}/{d#2}
}
\newcommand{\tdiffs}[2]%
{
  {d^2#1}/{d#2^2}
}
\newcommand{\tpdiff}[2]%
{
  {\partial#1}/{\partial#2}
}
\newcommand{\tpdiffs}[2]%
{
  {\partial^2#1}/{\partial#2^2}
}
\newcommand{\tpdiffx}[3]%
{
  {\partial^2#1}/{\partial#2 \partial#3}
}

\def\beq{\begin{equation}}
\def\eeq{\end{equation}}
\def\beqar{\begin{eqnarray}}
\def\eeqar{\end{eqnarray}}
\def\nn{\nonumber}

\def\gsim{{\lower.3em\hbox{${\>\buildrel > \over \sim\>}$}}}
\def\lsim{{\lower.3em\hbox{${\>\buildrel < \over \sim\>}$}}}

\newcommand{\vect}[1]%
{
  {\bf #1}
}

\def\b0{\vec{b}_{0}}

\def\grad{\nabla}

\def\d0par{\partial^0_{||}}
\def\hd0par{\hat{\partial}^0_{||}}
\def\1{\perp 1}
\def\2{\perp 2}

\newcommand\usecolor{black}

\begin{document}

\title{
  Modeling of convective cells, turbulence, and transport
  induced by \\
  a radio-frequency antenna in the tokamak boundary plasma
}

\author{M. V. Umansky}
\email{umansky1@llnl.gov}
\affiliation{
  Lawrence Livermore National Laboratory, Livermore, CA 94550, USA
}

\author{B. D. Dudson}
\affiliation{
  Lawrence Livermore National Laboratory, Livermore, CA 94550, USA
}

\author{T. G. Jenkins}
\affiliation{
  Tech-X Corporation, Boulder, CO 80303, USA  
}

\author{J. R. Myra}
\affiliation{
  Lodestar Research Corporation, Broomfield, CO 80023, USA
}

\author{D. N. Smithe}
\affiliation{
  Tech-X Corporation, Boulder, CO 80303, USA  
}

\date{\today}
\begin{abstract}

    The edge turbulence model Hermes (Dudson et al., 2017 Plasma
    Phys. Control. Fusion 59 05401) is set up for plasma boundary
    simulations with an RF antenna, using parameters characteristic of
    a tokamak edge. Cartesian slab geometry is used with thin plate
    limiters representing the ICRF antenna side-wall limiters.  Ad-hoc
    DC electric biasing of the limiters, motivated by calculations
    with VSim (Nieter et al., J. Comput. Phys. 196, 448 (2004)), represents an
    induced RF sheath rectified potential in the plasma turbulence
    model. Flux-driven turbulence simulations demonstrate a realistic
    distribution of plasma profiles and fluctuations. There is a clear
    effect of the antenna sheath voltage leading to formation of
    convective cells; bias-induced convective transport flattens the
    SOL density profile and fluctuations penetrate into the shadow
    region of the limiters as the bias voltage increases. Turbulent
    transport for impurity ions is inferred by following ion
    trajectories in the simulated plasma turbulence fields, showing
    Bohm-like effective diffusion rates. All in all, the model
    elucidates the key physical phenomena governing the effects of
    ICRF-induced antenna biasing on the tokamak boundary plasma.

\end{abstract}

\maketitle

\newpage

\section{Introduction}

For tokamak based fusion, plasma heating and current drive are
necessary, and radiofrequency (RF) heating and current drive have long
been recognized as essential tools for realizing a steady state
tokamak \cite{Kikuchi2012}.

The physics of ion cyclotron range of frequency (ICRF) wave
interaction with tokamak plasma is very rich; it includes RF-sheath
wall erosion, impurity production and transport, RF propagation and
power deposition in the SOL, edge, and hot core region. The conditions
of the scrape-off layer (SOL) plasma are known to affect the
efficiency of RF heating and current drive, and detailed understanding
of these effects is necessary to make accurate projections for future
devices. In particular, RF-induced convective cells may affect
plasma-material interactions. They may also affect the efficiency of
coupling between the RF antenna and the core plasma and are further
believed to be a cause of enhanced impurity transport
\cite{Noterdaeme_1993,Myra2021,Colas_2021,Zhang_2022,Parisot_2003}.

Convective cells are conceptually understood to be caused by ExB
drifts resulting from the spatial variation of the rectified RF sheath
\cite{DIppolito1998}. They manifest themselves in altering edge plasma
density profiles, also in enhanced impurity ion transport in the edge
\cite{DIppolito1998}.

In the present study, a plasma model is implemented to capture the
physics of RF convective cells and boundary plasma turbulence and
transport.



\section{Physics model}

\subsection{Hermes plasma model}
The physics model used in the present study is based on drift-fluid
equations implemented in the plasma turbulence code Hermes
\cite{Dudson2017,Dudson_2024}. This includes dynamic equations for
plasma density, potential vorticity, parallel electron momentum,
parallel ion momentum, electron pressure, and ion pressure as
described in detail in the Appendix.

\subsection{Simulation setup}
For the present study, the simulations are set up in a domain
representing a flux tube in the edge plasma, partially intercepted by
limiter plates orthogonal to the magnetic field, see
Fig. \ref{fig_fluxtube}.

Including geometric details of a realistic RF antenna is beyond the
scope of the present study; instead, a straight slab model is used to
represent the flux tube, see Fig. \ref{fig_simulation_geometry}.

In this geometry, the $x$ coordinate is radial, the $y$ coordinate is
parallel to the magnetic field, and the $z$ coordinate is
binormal. The $z$ coodinate is periodic, Neumann boundary conditions
are used on radial boundaries for the electric potential, density and
pressure, and sheath boundary conditions are implemented in y on the
end plates and on the limiter plates.

The dimensions of the domain are
taken 0.1 m in $x$, 20 m in $y$, and 1 m in $z$; the limiter plates
are located at $y$=6.53 m and $y$=8.21 m, and extend radially from the
outer radial boundary to the mid-point $x$=0.05 m.
\textcolor{\usecolor}{
  The size of the grid used in the described simulations is $N_x$=68,
  $N_y$=96, $N_z$=243.
}
The magnetic field is uniform, B=1 T, however a magnetic curvature
term is included in the model, with the effective curvature radius
taken $R_c$=1 m.

Hermes is a ``flux-driven'' model, which means that a source of plasma
density is used to calculate self-consistently both the plasma
turbulence and the average plasma profiles.

In the present simulations, the plasma density source has the spatial
form

\beq
S_n(x,y) = S_0 \exp
\left[
-\left(\frac{x-x_0}{L_{sx}}\right)^2
-\left(\frac{y-y_0}{L_{sy}}\right)^2
\right]
\label{e_dens_src}
\eeq

where the source center is at $x_0$=0.3 m, $y_0$=7.35 m, the scale
length factors are $L_{sx}$=0.1 m and
\textcolor{\usecolor}{
  $L_{sy}$=0.2 m
}
and, and $S_0$ is the
overall scaling factor chosen to produce realistic radial plasma
density profile. In the simulations discussed here, the plasma source
$S_n$ is set to create electrons and ions with temperature 20 eV.
\textcolor{\usecolor}{
  In the present simulations, the plasma density source is first set
  for a reference simulation case without the bias voltage, and the
  same source is used with a finite bias to determine the net effect
  of the bias.
}
\textcolor{\usecolor}{
  Note that in the described Hermes setup, despite the Neumann
  boundary conditions used on the radial boundaries, the dynamic
  fields are uniquely defined because the sheath boundary conditions
  on the poloidal boundaries provide a sink to balance the fixed
  plasma density source.
}

\subsection{Sheath boundary conditions}
The RF physics enters the turbulence model through the boundary
condition for the parallel electric current on the material surface,
Eq. (\ref{eq_bc_jpar}), where the electric potential $\phi_{wall}$
includes the effect of rectified sheath \cite{Myra2021} induced by the
RF antenna field. For the present simulations, an ad-hoc model is used
for rectified sheath potential $\phi_{rsh}$,

\beq
\phi_{rsh} = V_0 f(z) \exp \left( {\frac{x-x_{mid}}{L_x}} \right),
\label{eq_rect_sheath}
\eeq

Here $V_0$ is the scaling factor, in the simulations results shown
here $V_0$=0 (reference case), 50 V, and 100 V were used. The
parameter $x_{mid}$ corresponds to the radial mid-point,
$x_{mid}$=0.05 m, and $L_x$ is the spatial decay length taken as
$L_x$=0.01 m. The function $f(z)$ is the envelope function in the
periodic binormal (toroidal) coordinate, taken as



\beq
f(z) =
\begin{cases} 
      0:  & z  < 1/3 \\
      1:  & 1/3  \leq z \leq 2/3 \\
      0:  & z > 2/3 
   \end{cases}
\eeq

Since the $z$ coordinate is periodic, the chosen toroidal envelope
function $f(z)$ accounts for the fact that the RF antenna has a
limited toroidal span. The exponential factor in
Eq. (\ref{eq_rect_sheath}) enforces strong radial localization of the
rectified sheath potential at the leading edges of the limiters; this
choice is motivated by the form of the rectified sheath potential on
limiter surfaces found in detailed RF simulations \cite{Tierens_2024}.



\section{Simulations results}

\subsection{Radial profiles}

As a flux driven turbulence code, Hermes uses the provided sources of
particles and energy to establish the average plasma profiles.
Starting from a small perturbation seed, the simulation goes through
growth of linear instabilities that result in transport and profile
evolution, which comes to a nonlinear saturated state, and this
nonlinear solution is what we call here “saturated turbulence”.

Examining the mean plasma density and electron temperature profiles in
the saturated turbulence state, using the time average over the
turbulence fluctuations, one can establish that the simulation is
relevant to far-SOL plasma in a typical tokamak, with the densities on
the order of 10$^{18}$ m$^{-3}$ and temperatures 5-10 eV, and the
radial scale length on the order of 1 cm, see
Fig. \ref{fig_ni_te_mean}. Furthermore, comparing in the Figure the
radial density profiles for the reference case and for the 50 V bias
case, one can see that the average density profile is affected by the
RF sheath bias, which apparently flattens the radial density
profile, qualitatively similar to some experimental
results in TFTR \cite{DIppolito1998}.
\textcolor{\usecolor}{
  On the other hand, in some other experimental results the RF antenna
  leads to steepening of the radial plasma density profile which is
  attributed to $E{\times}B$ sheared flow suppression of plasma
  turbulence
  \cite{Weynants2005,Sun2013,Antar2010,Diab2024}. Furthermore, for the
  same experimental device, the effect of RF antenna can be both
  steepening and flattening of the radial plasma profiles, depending
  on the RF antenna polarity, as shown in Fig. 5 in the earlier cited
  paper by D\textsc{\char13}Ippolito et al. \cite{DIppolito1998}. Note
  that from the theoretical point of view, $E{\times}B$ shearing can
  both suppress and drive instabilities. The Kelvin-Helmholtz (KH)
  instability driven by sheared ExB flows was previously discussed in
  detail \cite{Popovich2010a} in the context of a physics model
  similar to that in Hermes (essentially a subset of Hermes
  equations).  It will require a more detailed investigation to
  resolve the apparent disagreement with the experimental evidence
  pointing to edge transport suppression by RF-induced sheared
  flows. However, one can speculate that the rate of flow shear in the
  present simulations ($\sim$10$^5$ 1/s, based on
  Fig. \ref{fig_ni_te_mean}) is not sufficient for suppressing edge
  turbulence. Note, that there is a power threshold for RF edge
  turbulence suppression in the experiments \cite{Diab2024}.
}

\subsection{Time slices of turbulent perturbations}

Next, in the saturated turbulence stage, we examine the electric
potential $\phi$ perturbation in the y-z plane perpendicular to the
magnetic field, for three different values of x: 0.04 m, 0.05 m, and
0.06 m. These correspond to the radial location in the main SOL, right
at the limiter tips, and in the private SOL between the limiters; the
third one includes the ``shadow'' region between the limiters. For the
``zero bias'' reference case, the results are shown in
Fig. \ref{fig_phi_yz_0bias}, and for the case with the bias voltage
scale factor $V_0$=50 V the results are shown in
Fig. \ref{fig_phi_yz_50bias}. We see that the $\phi$ perturbations are
aligned with the magnetic field, and we see that a shadow develops
between the limiters. Also, one can see that in the case with $V_0$=50
V, the bias voltage is added to the electric potential
perturbation. Note the change in scale on the color paletts.

Next, examining the plasma \textcolor{\usecolor}{electric potential}
perturbations, using the
cross-section in the x-z plane in Fig. \ref{fig_phi_xz}, elongated
"blobs" are observed.  These structures are roughly circular in
physical space; the elongation is due to the 10:1 ratio of scales on
the two axes. Comparing the reference case with $V_0$=0 and the case
$V_0$=50 V, one can observe some interesting features in the biased
case. First, the penetration of fluctuations beyond antenna limiters
(towards the radial wall at x = 0.10) is stronger with increased
bias. Increasing the bias voltage to 100 V confirms the trend, for
both the penetration of fluctuations into the shadow region and for
the convective cell formation.
\textcolor{\usecolor}{
  Similar conclusion can be drawn from examining the spatial
  distribution of plasma density RMS perturbations (not shown in the
  plots); the penetration of fluctuations into the shadow region grows
  with the bias voltage.
}
\textcolor{\usecolor}{
  While the zero-bias case has only regular density perturbations
  (density filaments), the biased cases also have larger scale details
  which can be attributed to a convective cell emerging, as shown in
  Fig.  \ref{fig_ni_xz_swirl} where a set of slices of plasma density
  perturbation is shown for nine different y-locations, uniformly
  distributed along the y coordinate. One can see that a swirl-like
  feature is formed close to ${z}$=0.7, especially visible for
  intermediate y corresponding to the locations between the limiter
  plates.
}

\subsection{Impurity ion transport}

One of the most significant manifestations of the RF physics
interference with tokamak edge turbulence is the effect on impurity
ions.
\textcolor{\usecolor}{
  It is understood that the RF sheath leads to larger sputtering rates
  of the impurity ions due to the increased sheath potentials, and the
  induced convective cells lead to larger plasma fluxes to the
  material surfaces \cite{DIppolito1998}. Furthermore, the induced
  convective cells may lead to enhanced transport of sputtered
  impurity ions to the main plasma.
}

In the present study, the impurity ion transport is examined by
calculating drift orbits of passively advected impurity particles in
the simulated turbulent fields.

A set of Monte-Carlo particles is used, representing tungsten ions,
launched at the same $x_0, y_0$ locations corresponding to the limiter
tips radially and the mid-point between the limiters poloidally, and
uniformly distributed $z_0$, with the initial $V_{||,0}$ randomly
drawn from the Maxwellian distribution corresponding to the plasma
temperature $T_i=$5 eV. The standard guiding center equations
are used, greatly reduced since the magnetic field is uniform,

\beqar
\frac{d}{dt} \vec{r} = V_{||} \hat{b} + \vec{V}_E \\ \nn
\vec{V}_E = \frac{c}{B^2} [E \times B]
\eeqar

The result of this calculation is shown in Fig. \ref{fig_mc_traj}
where the trajectories of the Monte-Carlo particles are shown for the
reference case and for the 50 V bias case.

In both cases, one can infer from the Monte-Carlo particles
trajectories the effective diffusion coefficient on the order of Bohm
diffusion, D $\sim$ 5 $m^2/s$, from the trajectories
spreading. Furthermore, for the biased case one can also infer a
convective ``inward pinch'' term, V $\sim$ 200 m/s,
\textcolor{\usecolor}{
  while for the non-biased case the mean convective velocity is insignificant.
}
More detailed analysis of the Monte-Carlo particles trajectories shows
that in the biased case the particles are caught in the swirl
corresponding to the convective cell discussed earlier.



\section{Summary and conclusions}

The BOUT++ based edge turbulence model Hermes is set up for
simulations with RF biased antenna limiters, with parameters
characteristic of tokamak edge, using Cartesian slab geometry with
thin plate limiters. RF field effects enter the model via ad-hoc
electric biasing of the limiters representing the induced rectified RF
sheath. The model reproduces the basic phenomenology of RF effects on
the edge plasma, showing realistic distribution of plasma profiles and
fluctuations, and the effects of the antenna sheath on plasma profiles
and fluctuations.  In the simulations, bias-induced convective
transport flattens SOL density profile, which is consistent with
existing experimental observations. Furthermore, edge plasma
turbulence and the convective cell strongly affect the transport of
impurity ions, resulting in effective anomalous diffusion and
convection.

For the next steps, it is envisioned to include in the model a direct
calculation of the RF sheath bias voltage, e.g., based on the VSim
code, to provide a self-consistent description of the RF antenna and
edge plasma dynamics.
\textcolor{\usecolor}{
Thus, the present model, including the use of an
ad-hoc bias electric potential, sets the stage for a more accurate and
detailed description of edge plasma physics coupled with an RF
antenna.
}
\textcolor{\usecolor}{
However, even at the level of the present model (where the RF bias and
plasma dynamics are not self-consistent) there are standing issues
that deserve further analysis, e.g., understanding the effects of
RF-induced ExB flow shear on turbulence suppression vs. exciting the
Kelvin-Helmholtz instabilities. Due to the apparent complexity of the
interplay between edge plasma turbulence and RF-induced sheared E×B
flows, a first-principles edge plasma model, such as Hermes, is
necessary for uncovering the underlying physics of coupling between
edge plasma turbulence, transport, and RF antenna field.
}


\section{Appendix: Hermes model summary}

\renewcommand{\theequation}{A-\arabic{equation}} 
\setcounter{equation}{0}


Electron density
\beq
\pdiff{n}{t} = - \grad \cdot [n
  ( \vec{v}_{E \times B} + \vec{b} v_{||e} + \vec{v}_{de})
] + S_n
\eeq

Parallel electron momentum
\beq
\pdiff{m_e n v_{||e}}{t} = - \grad \cdot [ m_e n V_{||e}
  ( \vec{v}_{E \times B} + \vec{b} v_{||e} + \vec{v}_{de})
] - \vec{b} \cdot \grad p_e  - e n E_{||} + F_{ei}
\eeq

Electron pressure
\beqar
3/2 \pdiff{p_e}{t} = - \grad \cdot
[
  \frac{3}{2} p_e (\vec{v}_{E \times B} + \vec{b} v_{||e}) + \frac{5}{2} p_e \vec{v}_{de}] -
p_e \grad \cdot (\vec{v}_{E \times B} + \nn \\
\vec{b} v_{||e})
\grad \cdot (\kappa_{||e} b b \grad \cdot T_e) + S_{Ee} + W_{ei}
\eeqar

Parallel ion momentum
\beqar
\pdiff{m_i n v_{||i}}{t} = - \grad \cdot [ m_i n V_{||i}
  ( \vec{v}_{E \times B} + \vec{b} v_{||i} + \vec{v}_{di})
] - \vec{b} \cdot \grad p_i  + Z_i e n E_{||} - F_{ei}
\eeqar

Ion pressure
\beqar
3/2 \pdiff{p_i}{t} = - \grad \cdot
[
  \frac{3}{2} p_i (\vec{v}_{E \times B} + \vec{b} v_{||i}) + \frac{5}{2} p_i \vec{v}_{di}] -
p_i \grad \cdot (\vec{v}_{E \times B} + \nn \\
\vec{b} v_{||i})
\grad \cdot (\kappa_{||i} b b \grad \cdot T_i) + S_{Ei} + S_n \frac{1}{2} m_i n_i v_{||i}^2
- W_{ei}
\eeqar

Potential vorticity
\beqar
\pdiff{\varpi}{t} = -
\grad \cdot \left(
\frac{m_i}{2 B^2} \grad_{\perp} (v_{ExB} \cdot \grad p_i) + \frac{1}{2} \varpi \vec{v}_{ExB} 
\right)
+ \frac{m_i n_0}{2 B^2} \grad_{\perp}^2 \left( \vec{v}_{ExB} + \frac{b}{n_0 B} \times \grad p_i \right)
\nn \\
+ \grad \cdot \left( j_{||} \right)
+ \grad \cdot \left( (p_e + p_i) \grad \times \frac{b}{B} \right)
\eeqar

where the vorticity $\varpi$ is defined as

\beq
\varpi = - \grad \cdot
\left[
\frac{m_i n_0}{B^2} \grad_{\perp} \left( \phi + \frac{p_i}{n_0} \right)
\right]
\eeq

\medskip
The dynamic equations are supplemented with the sheath boundary
conditions on material surfaces, as follows.
\medskip

The parallel velocity condition (Bohm condition),
\beq
V_{||i} = C_s
\eeq

The heat flux boundary condition,
\beq
q_{||e,i} = \delta_{e,i} N_i C_s T_{e,i}
\eeq

The parallel current condition,

\beq
j_{||} = e N_i C_s
\left[
1 - \sqrt{\frac{1}{2\pi} \frac{M_i}{m_e}} \exp{(-
  (\phi-\phi_{wall})
  /T_e)}
\right]
\label{eq_bc_jpar}
\eeq

Here $C_s$ is the sound speed, $\delta_{e,i}$ are the sheath heat
transmission coefficients calculated from the Tskhakaya and Kuhn
model \cite{}.

\section*{Acknowledgements}

This work was performed under the auspices of the U.S. Department of
Energy by Lawrence Livermore National Security, LLC, Lawrence
Livermore National Laboratory under Contract DE-AC52-07NA27344, and
supported by the U.S. DOE Office of Science, Office of Fusion Energy
Sciences, and Office of Advanced Scientific Computing Research through
the Scientific Discovery through Advanced Computing (SciDAC) project
Center for Advanced Simulation of RF - Plasma - Material Interactions.

\clearpage
\newpage
\medskip
\medskip
\medskip
\bibliography{rfscidac_refs}

\clearpage
\clearpage

\section{Figure captions}

\begin{figure}[h] 
\caption{
The flux-tube domain used in the model. The vertical planes represent
the side-walls of the RF antenna enclosure.
\hfill
}
\label{fig_fluxtube}
\end{figure}

\begin{figure}[h] 
\caption{
Cartesian geometry used in the simulations. The leading edges of the
limiter plates (shown in red) are biased due to the rectified sheath
potential, as discussed in the text.
\hfill
}
\label{fig_simulation_geometry}
\end{figure}

\begin{figure}[h] 
\caption{
Mean plasma density (top) and electron temperature (bottom) profiles
averaged over the turbulent fluctuations for the reference case and
for the 50 V bias case. In the biased case it is apparent that the
density profile is flattened.
\hfill
}
\label{fig_ni_te_mean}
\end{figure}

\begin{figure}[h] 
\caption{
Electric potential perturbations for the zero-bias case. The
filamentary structure of perturbations is apparent, along with the
effect of the gap in the shadow region between the limiters.
\hfill
}
\label{fig_phi_yz_0bias}
\end{figure}

\begin{figure}[h] 
\caption{
Electric potential perturbations for the $V_0$=50 V case. Compared to
the zero-bias case in Fig. (\ref{fig_phi_yz_0bias}), here one can
observe the added DC electric potential due to the bias voltage.
\hfill
}
\label{fig_phi_yz_50bias}
\end{figure}

\begin{figure}[h] 
\caption{
\textcolor{\usecolor}{
Electric potential perturbations for representative time slices for
the zero-bias reference case (top left), the $V_0$=50 V case (bottom
left), and $V_0$=100 V case (bottom right). It is apparent that the
fluctuations penetrate deeper beyond the limiters tip level (dashed
line) for larger bias voltage.  }
\hfill
}
\label{fig_phi_xz}
\end{figure}

\begin{figure}[h] 
\caption{
Plasma density perturbations for the $V_0$=50 V case, shown at nine
y-locations uniformly distributed over the domain length. Emergence of
convective cells is apparent near z=1/3 and z=2/3 where the bias
potential changes abruptly, according to Eq. (\ref{eq_rect_sheath}).
\hfill
}
\label{fig_ni_xz_swirl}
\end{figure}

\begin{figure}[h] 
\caption{
Monte-Carlo particles trajectories in the turbulent fluctuations
fields, for the zero-bias case (top) and 50 V bias case (bottom). In
both cases, the radial spreading of the particles can be characterized
by a Bohm-like turbulence diffusion coefficient D$\sim$ 5 m$^2$/s; in
the biased case there is also an average radial drift which can be
characterized by a radial inward pinch velocity $V \sim$ 200 m/s.
\hfill
}
\label{fig_mc_traj}
\end{figure}

\clearpage



{
\LARGE{Fig. 1}
\begin{figure}
\includegraphics[scale=0.5]{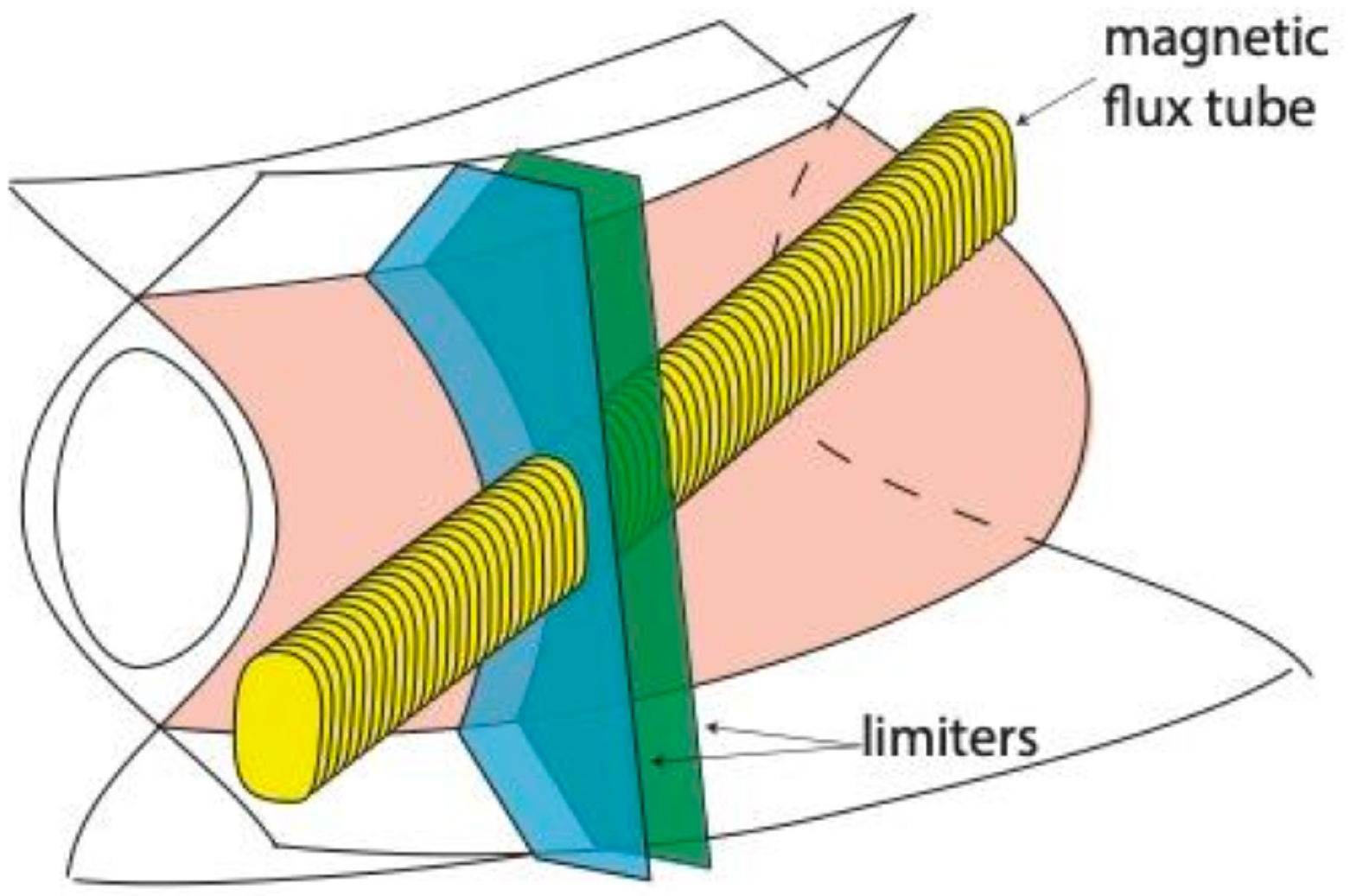}
\end{figure}
\clearpage
}

{
\LARGE{Fig. 2}
\begin{figure}
\includegraphics[scale=0.5]{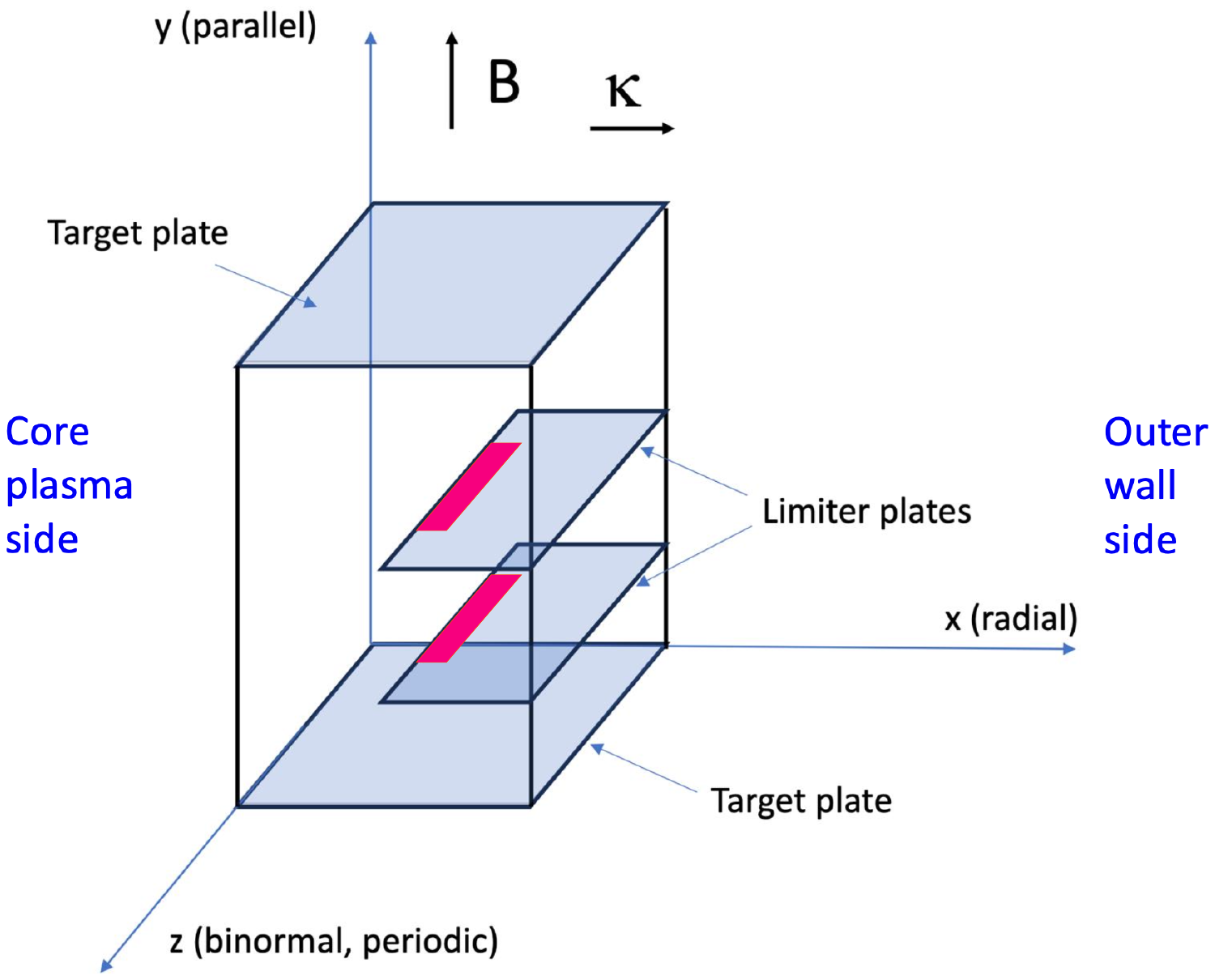}
\end{figure}
\clearpage
}

{
\LARGE{Fig. 3}
\begin{figure}
\includegraphics[scale=0.4]{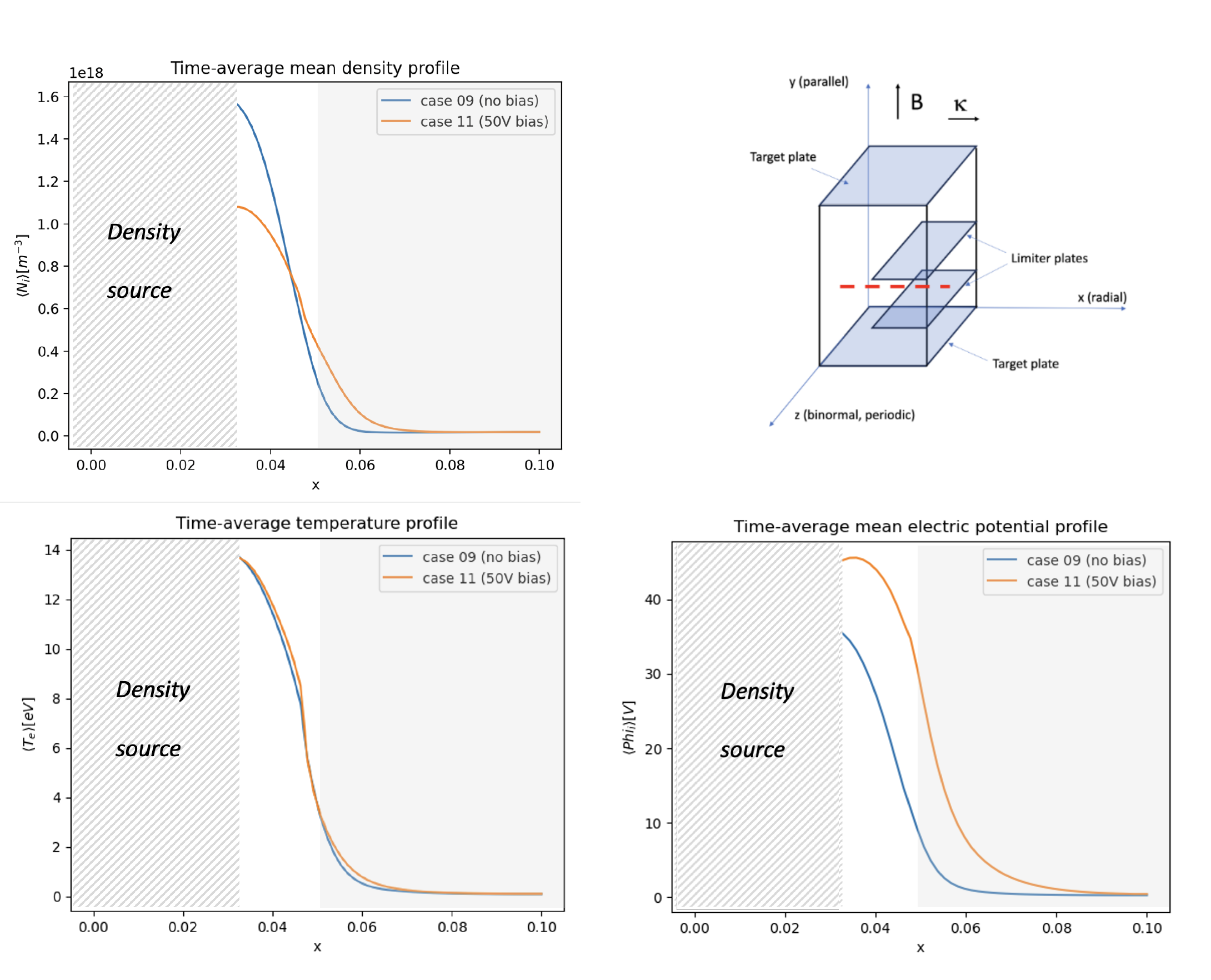}
\end{figure}
\clearpage
}

{
\LARGE{Fig. 4}
\begin{figure}
\includegraphics[scale=0.5]{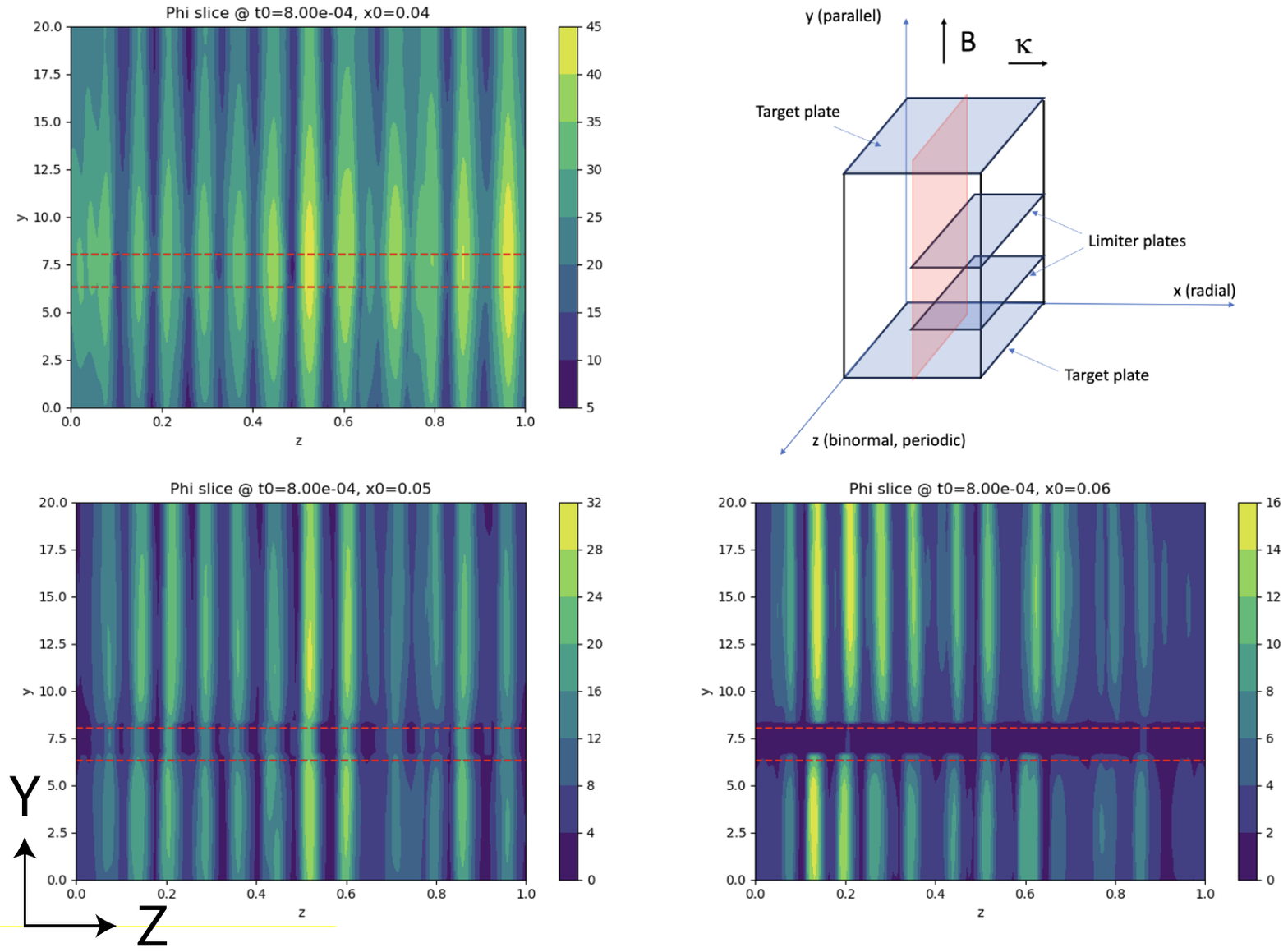}
\end{figure}
\clearpage
}

{
\LARGE{Fig. 5}
\begin{figure}
\includegraphics[scale=0.5]{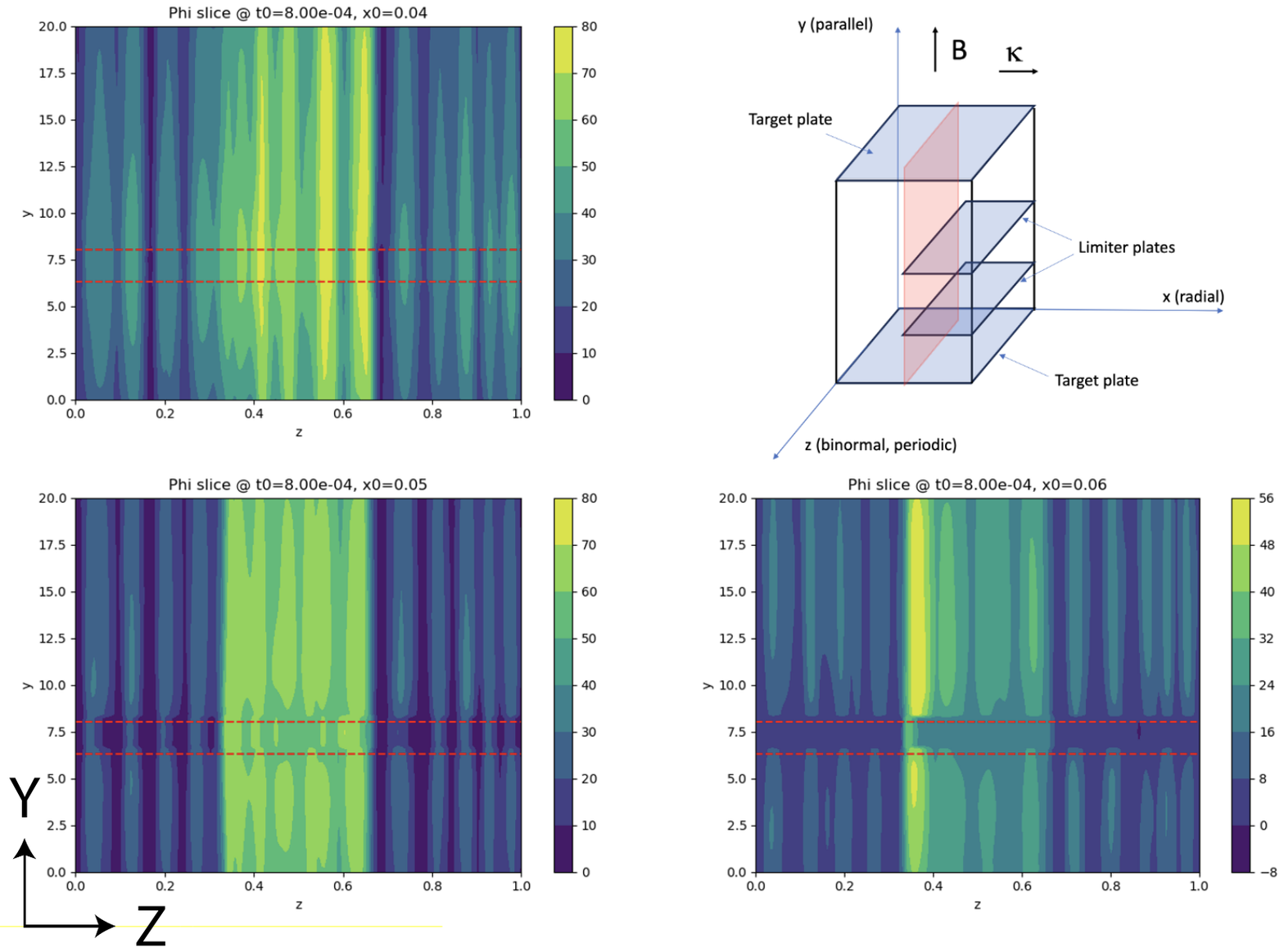}
\end{figure}
\clearpage
}

{
\LARGE{Fig. 6}
\begin{figure}
\includegraphics[scale=0.6]{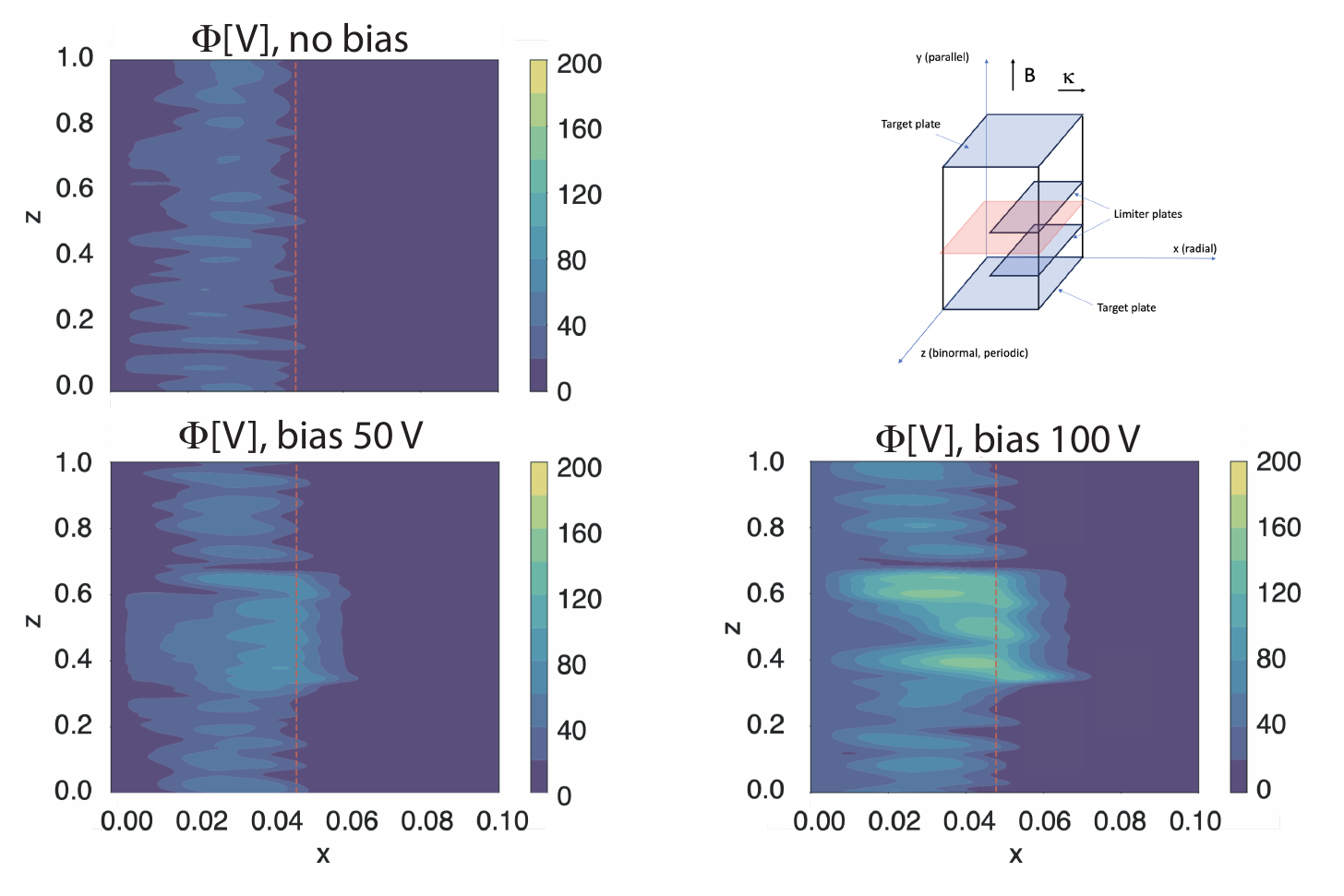}
\end{figure}
\clearpage
}

{
\LARGE{Fig. 7}
\begin{figure}
\includegraphics[scale=0.6]{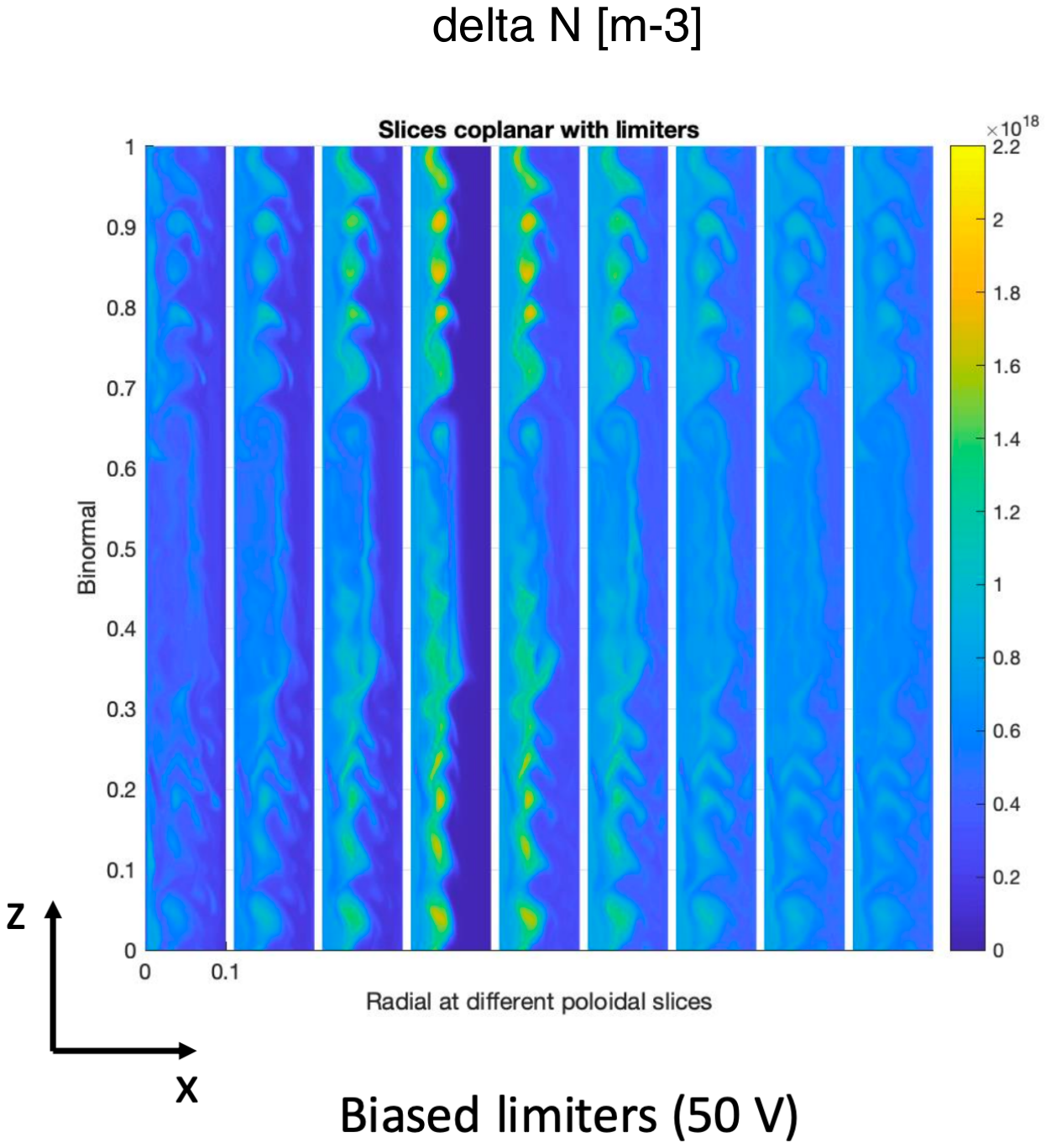}
\end{figure}
\clearpage
}

{
\LARGE{Fig. 8}
\begin{figure}
\includegraphics[scale=0.75]{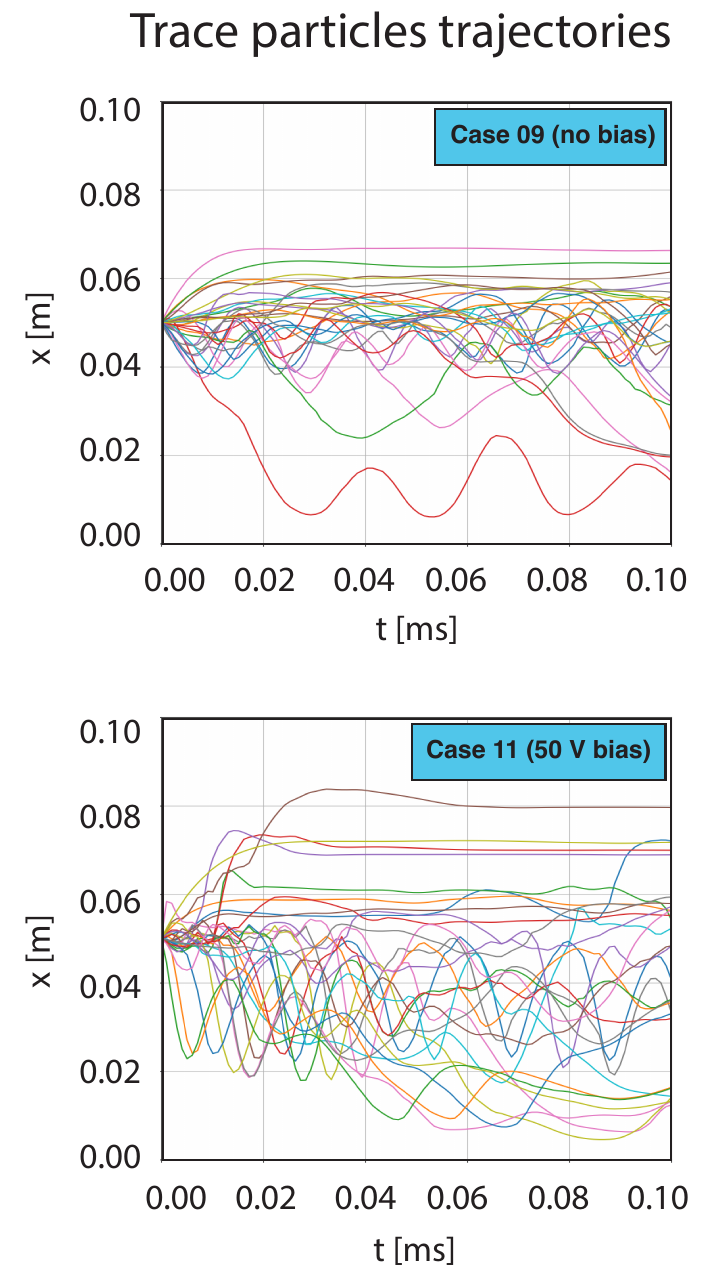}
\end{figure}
\clearpage
}


%
%
\end{document}